\documentstyle[aps,prl,tabularx,graphicx,twocolumn]{revtex}

\def\eqref#1{(\ref{#1})}

\begin{document}

\title { Polyelectrolytes in the presence of
multivalent ions: gelation versus segregation }

\author{ A. V. Ermoshkin, M. Olvera de la Cruz}

\address{Department of Materials Science and Engineering,\\
Northwestern University, Evanston, IL, USA 60208-3108}

\date{\today}

\maketitle

\begin{abstract}
We analyze solutions of strongly charged chains bridged by linkers
such as multivalent ions. The gelation induced by the strong
short range electrostatic attractions is dramatically suppressed
by the long range electrostatic correlations due to the charge
along the uncrosslinked monomers and ions. A modified
Debye-H$\rm\ddot u$ckel approach of crosslinked clusters of
charged chains is used to determined the mean field gelation
transition self-consistently. Highly dilute polyelectrolyte
solutions tend to segregate macroscopically. Semidilute solutions
can form gels if the Bjerrum length $l_B$ and the distance between
neighboring charged monomers along the chain $b$ are both greater
than the ion size $a$.

\end{abstract}

\pacs{}

Linear polyelectrolytes are ubiquitous in biology given that
nucleic acids and most proteins are charged. Moreover, they have
important technological applications as gelling and drag reduction
agents. Long linear polyelectrolytes are typically water soluble
in low ionic strength monovalent salt solutions due to a net
repulsion between the charged monomers. Linking agents such as
multivalent ions of valence $z>1$, however, modify the stability
of polyelectrolytes aqueous solutions
\cite{Widom80,Delsanti95,Joanny95,Raspaud98,Tagabi}. The stability
of biopolymers in various ionic media has important implications
in biotechnological processes \cite{Sikorav}.

The solubility of linear polyelectrolyte in multivalent salts has
been extensively studied in dilute solutions of continuously
charged flexible chain \cite{Solis00} and related
systems\cite{Rouzina96}. Only few studies, however, have discussed
the stability of the chains in semidilute solutions
\cite{Delsanti95,Joanny95}. In particular, the strong binding
between small (dehydrated) multivalent metallic ions and charged
($O^-$) groups in different chains opens the possibility of a
gelation transition in semidilute
solutions\cite{Joanny80,Delsanti}. The competition of gelation and
phase separation has been recently studied in neutral chains with
linkers \cite{Kumar,Erukh02}. The charge along the chains,
however, will strongly modify the phase diagram. Here we describe
the gelation by multivalent ions including long and short range
correlation self-consistently in the analysis.

Our model treats the strong short range correlations driven by the
multivalent cations explicitly using bridges or crosslinks that
change the structure function of the system and therefore the long
range electrostatic correlations. The long range correlations are
accounted via a modified mean field approach known as the Random
Phase Approximation (RPA) on the linked system. RPA describes
properly long range correlations in polyelectrolyte solutions
without short range correlations\cite{Pedro}. Therefore, by using
as a reference state of RPA a strongly correlated system of chains
linked by multivalent ions, we can in principle determine
self-consistently the gelation transition in the system of charged
chains. We consider only flexible charged chains to avoid
difficulties found in the gel structures of charged rigid rods
\cite{Bruisma} due to the strong orientational dependent
electrostatic interactions\cite{Muthu,Borukhov}.

Consider a polymer solutions where every chain monomer carries a
negative charge. The concentration of monomers is $\rho_m$, and of
monovalent counterions is $\rho_m^c=\rho_m$, and we assume for
simplicity that all of them are completely dissociated from the
chains. We denote the concentrations of multivalent ions and their
counterions as $\rho_i$  and $\rho_i^c$, respectively
($\rho_i^c=z\rho_i$). Let us assume that $z=2$. The interaction
between the multivalent ions and the chains, in principle, leads
to the formation of monocomplexes ($mc$) and dicomplexes ($dc$). A
$mc$ is formed when an ion binds onto one negative site of the
polyelectrolyte. Formation of a $dc$ brings two negative sites
together. The competition between the formation of $mc$ and $dc$
is described by the process $2mc\longleftrightarrow dc +
free\,\,ion $. The equilibrium constant $p$ of this process can be
estimated as\cite{Joanny95} $p=K_2\exp(\kappa l_B)/K_1^2$, where
$l_B=e^2/\epsilon k_B T$ is the Bjerrum length and the inverse
screening length $\kappa$ is given by $\kappa^2=4\pi l_B\sum
z_\alpha^2\rho_\alpha$ with the sum taken over all dissociated
charges. $K_1$ and $K_2$ are the statistical weights of $mc$ and
$dc$ respectively. Assuming that all ions and the monomers of the
chains have the same size $a$ we can estimate these weights
following Fuoss\cite{Fuoss58}
\begin{eqnarray}
K_1&=&\frac{4\pi}{3}a^3\exp(l_B/a) \label{K_1} \\
K_2&=&\frac{1}{2}\left(\frac{4\pi}{3}a^3\right)^2\exp(3l_B/a) \label{K_2}
\end{eqnarray}
As shown in Ref. \cite{Joanny95} dicomplexation is favorable if
$p>1/12$. Since in our work we consider strongly charged chains
regimes with $p\sim10$, we neglect the possibility of forming
monocomplexes.

Dicomplexation processes may have two different scenarios. If the
two sites that form a dicomplex belong to the same chain then this
chain collapses onto itself. If these sites, however, are from
different chains then a crosslink is formed. The monomolecular
collapse is observed when the multivalent ions nearly neutralize
the chains at \cite{Solis00}
\begin{equation}
\rho_i^{\rm coll}=\frac{\rho_m}{z} \label{collapse}
\end{equation}
Here we consider the formation of branched structures by
crosslinks at $\rho_i < \rho_i^{\rm coll}$.

We write the free energy of the system in the form
\begin{equation}
F=F^*+F_{\rm ref}+F_{\rm el}
\label{F}
\end{equation}
$F^*$ accounts for excluded volume interactions and can include
small effective short range monomer attractions ($\chi \leq 2$) to
account for slightly hydrophobic chain backbones. It can be
approximated as
\begin{equation}
\frac{F^*a^3}{TV}=(1-\phi)\log(1-\phi) -\chi \phi_m^2
\label{Fstar}
\end{equation}
Here $T$ is the temperature in units of $k_B$, $V$ is the volume,
$\phi_m= a^3\rho_m$, and $\phi$ is the total volume fraction of
all components $\phi=a^3(\rho_m+\rho_m^c+\rho_i+\rho_i^c)$.
$F_{\rm ref}$ corresponds to free energy of the system without any
long range electrostatic interactions, the reference system,
\begin{equation}
\frac{F_{\rm ref}}{TV}=\sum\limits_{\{C\}} \rho_C\ln\frac{\rho_Cs_C}{ew_C}
\label{Fref}
\end{equation}
The sum is taken over all possible structures $C$ of clusters
formed due to crosslinking ($\{C\}$ also includes trivial clusters
such as free ions and counterions), and $\rho_C$, $w_C$ and $s_C$
represent concentration, statistical weight and symmetry index of
a cluster $C$, respectively.

The last term $F_{el}$ in Eq. \eqref{F} takes into account all
long range electrostatic effects and can be written in the form
\cite{Borue88}
\begin{equation}
\frac{F_{\rm el}}{TV}=
\frac{1}{2(2\pi)^3}\int\left[\ln\frac
{\det\|U^{ij}_{\bf k}+(g^{-1})^{ij}_{\bf k}\|}
{\det\|(g^{-1})^{ij}_{\bf k}\|}-
\sum\limits_i\rho_i U^{ii}_{\bf k}\right]d{\bf k}
\label{Fel}
\end{equation}
where the matrix of correlation functions
${\bf g_ k}=\|g^{ij}_{\bf k}\|$ has the following form
\begin{equation}
g^{ij}_{\bf k}=\sum\limits_C\rho_Cg(C)^{ij}_k
\label{gij}
\end{equation}
Here $g(C)^{ij}_{\bf k}$ is the scattering factor of a  cluster characterized by a
structure $C$
\begin{equation}
g(C)^{ij}_{\bf k}=\sum\limits_{m,n}\left<e^{i{\bf k}({\bf r}_i^m-{\bf r}_j^n)}\right>_C
\label{gCij}
\end{equation}
with $m$ and $n$ running over all $i$-type and $j$-type units in $C$ cluster
respectively.

We calculate free energy of the reference system assuming that all
complex clusters have only tree-like architectures
\cite{Joanny95,Erukh95,Semenov98}
\begin{equation}
F_{\rm ref}=F_{\rm ref}^{\rm id}+F_{\rm ref}^{\rm comb}+F_{\rm ref}^{\rm cross}
\label{Fref1}
\end{equation}
$F_{\rm ref}^{\rm id}$ is associated with all translational
entropies,
\begin{equation}
\frac{F_{\rm ref}^{\rm id}}{TV}=
\frac{\rho_m}{N}\ln\rho_m+
\rho_m^c\ln\rho_m^c+
\rho_i\ln\rho_i+
\rho_i^c\ln\rho_i^c
\label{Fid}
\end{equation}
$F_{\rm ref}^{\rm comb}$ is the term obtained from the number of
possibilities to choose the monomers and the ions that participate
in formation of crosslinks,
\begin{eqnarray}
\frac{F_{\rm ref}^{\rm comb}}{TV}=
\rho_m(\Gamma_m\ln\Gamma_m+(1-\Gamma_m)\ln(1-\Gamma_m))+ \nonumber \\
\rho_i(\Gamma_i\ln\Gamma_i+(1-\Gamma_i)\ln(1-\Gamma_i))
\label{Fcomb}
\end{eqnarray}
where $\Gamma_m$ and $\Gamma_i$ are the fractions of the monomers
and the ions respectively that belong to the crosslinks. Because
we neglect the formation of monocomlexes these fractions are
related by $\rho_m\Gamma_m=z\rho_i\Gamma_i$. The last term in Eq.
\eqref{Fref1} corresponds to the crosslinking free energy,
\begin{equation}
\frac{F_{\rm ref}^{\rm cross}}{TV}=-
\rho_m\Gamma_m\ln\frac{\rho_m\Gamma_m K_2^{1/2}}{e}
\label{Fcross}
\end{equation}

We use Eqs. \eqref{Fref1}-\eqref{Fcross} for the reference free
energy to  determine the instability of the system due to the
formation of the infinitely large network or gel. The gelation
line is identified by the divergence of the cluster's weight
average $N_w$. We note that $N_w\to\infty$ corresponds to a
singularity of the second derivatives of the reference free energy
\cite{Erukh95}
\begin{equation}
\det\left\| \frac{\partial^2 F_{\rm
ref}}{\partial\rho_\alpha\partial\rho_\beta} \right\|=0
\label{instability}
\end{equation}
Here the set $\{\rho_\alpha\}$ includes $\rho_i$, $\rho_m$ and
$\rho_m^{\rm cross}=\rho_m\Gamma_m$. Representing $F_{\rm ref}$
only as a function of these three concentrations and evaluating the
determinant \eqref{instability} we get the critical value
$\Gamma_m^*$ that corresponds to the formation of the network
\begin{equation}
\Gamma^*_m=\frac{1}{(z-1)(N-1)}
\label{Gamma_star}
\end{equation}

The last term of the free energy $F_{\rm el}$ given by Eq.
\eqref{Fel} is calculated as follows. First we evaluate the
$5\times5$ matrix of correlation functions $\bf g_k$ which has the
following components: $g_{11}=\rho_m^c$ is the concentration of
counterions dissociated from the polymer chains,
$g_{22}=\rho_i(1-\Gamma_i)$ is the concentration of free
multivalent ions, $g_{33}=\rho_i^{c}$ is the concentration of
counterions dissociated from the multivalent ions, and the
correlation functions between the connected units are denoted by
$g_{44}$, $g_{45}=g_{54}$ and $g_{55}$, where $4$ and $5$ stand
for non-crosslinked monomers and crosslinked aggregates
respectively. In the present work we neglect dipole-dipole and
dipole-charge interactions, therefore, $g_{45}$ and $g_{55}$ have
no effect on the free energy of the system. We calculate the
monomer-monomer correlations $g_{44}$ using the diagrammatic
technique described in Ref. \CITE{Ermosh99}. By introducing the
concentration $t$ of all branching structures that could be
attached to a chain monomer, and the concentration $\cal T$ of all
non-associated monomers, we obtain the total concentration of
monomers $\rho_m=t\cal T$, and the fraction of associated monomers
$\Gamma_m=1-1/t$. By finding the correlation function between two
monomers that belong to the same chain $\Sigma(k)=({\cal
T}/t)(g(k)-1)$ and taking into account all possible arrangements
of linear chains into tree-like structures we get
\begin{equation}
g_{44}(k)=\rho_m(1-\Gamma_m)\frac{1+(1-z\Gamma_m)(g(k)-1)}{1-\Gamma_m(z-1)(g(k)-1)}
\label{g44_final}
\end{equation}
where $g(k)$ is the structure factor of a chain defined as
\begin{equation}
g(k)=\frac{1}{N}\sum\limits_{i,j}\left<e^{-i{\bf k}({\bf r}_i-{\bf r}_j)}\right>
\label{gkdef}
\end{equation}
Strongly charged chains in semidilute solutions are stretched at
length scales of few monomers $n$ due to the locally unscreened
electrostatic repulsions. We assume they behave like rods on short
length scales, although simulations reveal slightly less stretched
local conformations\cite{Stevens95}. On larger scales the chains
obey Gaussian statistics with $N/n$ segments per chain. For $g(k)$
we use the approximated expression
\begin{equation}
g(k)-1=\frac{n-1}{1+nkb}+\frac{N-n}{1+Nnk^2b^2/12}
\label{gk}
\end{equation}
where $b$ is the average distance between the neighbor monomers
along the chain, which gives the right limits and it is reasonably
accurate. Though in ref. \cite{Stevens95} it is argue that $n$ is
nearly $\phi_m$ independent, we estimated it here
as\cite{Dobrynin99}
\begin{equation}
n=\frac{1}{(\phi_mb^3)^{1/2}}
\label{n}
\end{equation}
to obtain a bound on the gelation in the largest possible
perturbed chain conformation. We also calculate here the gelation
for the least perturbed chain conformations, Gaussian chains, by
putting $n=1$ in Eq. \eqref{gk}; ideal chain statistics are
observed in charged chains in certain hydrophobic backbone regimes
\cite{Holm}.

When the fraction of associated monomers $\Gamma_m$ reaches the
critical value $\Gamma_m^*$ given by Eq. \eqref{Gamma_star} then
the correlation function $g_{44}(k)$ diverges at $k=0$. This
divergence represents correlations on the infinite scale and, as
discussed above, corresponds to the network formation.

To obtain the long range electrostatic contribution to the free
energy given by Eq. \eqref{Fel} we introduce the following
interaction potential between the charges \cite{Sasha}
\begin{equation}
U^{ij}(r)=z_iz_jl_B\frac{1-e^{-r/a}}{r}
\label{Ur}
\end{equation}
which allows us to account for short range repulsions due to the
hard-sphere nature of the charges.

Taking the Fourier transform of $U^{ij}(r)$ we write the
electrostatic free energy \eqref{Fel} in the form
\begin{eqnarray}
\frac{F_{\rm
el}}{TV}=\frac{1}{4\pi^2}\int\biggl[\ln\left(1+U(k)(g_{44}(k)+\kappa^2)\right)
\nonumber \\
-U(k)(\rho_m(1-\Gamma_m)+\kappa^2)\biggr]k^2dk
\label{Fel_int}
\end{eqnarray}
where $\kappa^2=\rho_m+z\rho_i+z^2\rho_i(1-\Gamma_i)$ and
\begin{equation}
U(k)=\frac{4\pi l_B}{k^2(1+a^2k^2)}
\label{Uk}
\end{equation}

\begin{figure}
\includegraphics[width=8cm]{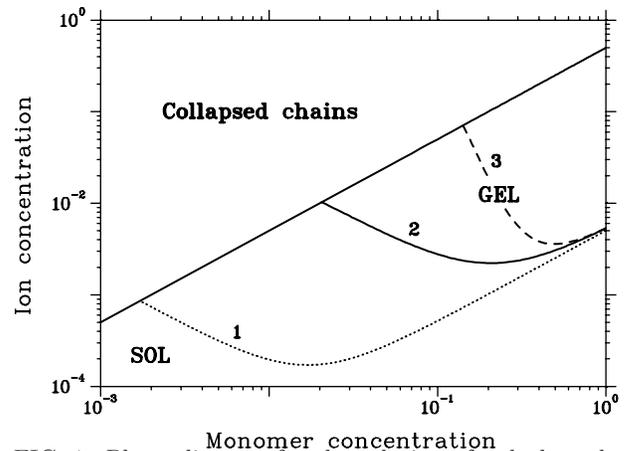}
\caption{Phase diagram for the solution of polyelectrolyte chains
in the presense of divalent ions. Concentrations are given in
$a^3$ units, $l_B/b=2$, $b=a$, $N=100$.} \label{portret}
\end{figure}

We obtain the fraction of crosslinked monomers in the system
$\Gamma_m^{\rm min}$, for given $\rho_m$, $\rho_i$ and $l_B/b$ by
minimizing the total free energy with respect to $\Gamma_m$.
Because the correlation function $g_{44}$(0) diverges at
$\Gamma_m=\Gamma_m^*$ we find a minimum of the free energy in the
interval $\Gamma_m\in[\,0,\Gamma_m^*]$. If the minimum is reached
inside this interval then the system contains only finite size
clusters. If the minimum is reached at $\Gamma_m^{\rm
min}=\Gamma_m^*$ then an infinitely large network is formed in the
system. By numerically solving
\begin{equation}
\Gamma_m^{\rm min}(\rho_i^*,\rho_m,K(\xi))=\Gamma_m^*
\label{gelation_min}
\end{equation}
we obtain the gelation line $\rho_i^*(\rho_m)$. For a given
$\rho_m$, if $\rho_i<\rho_i^*$ the system contains only finite
clusters, if $\rho_i\ge\rho_i^*$ an infinitely large network or
gel is formed. The results are shown in Figure \ref{portret}.
Above the straight line which corresponds to Eq. \eqref{collapse}
the chains are collapsed. The dash line labelled as $1$ is the
gelation line when the long range electrostatic interactions are
not included (i.e., setting $F_{el}=0$ in Eq. \eqref{Fel_int}),
and it is independent of the chain structure function in the mean
field model . Line $2$ is the gelation transition obtained for
locally stretched chains (the structure factor given by Eq.
\eqref{gk}), and line $3$ is the gelation obtained for Gaussian
chain statistics at all length scales ($n=1$). The gelation
transition changes by orders of magnitude and depends on the chain
conformation when long range electrostatic interactions are
included.

\begin{figure}
\includegraphics[width=8cm]{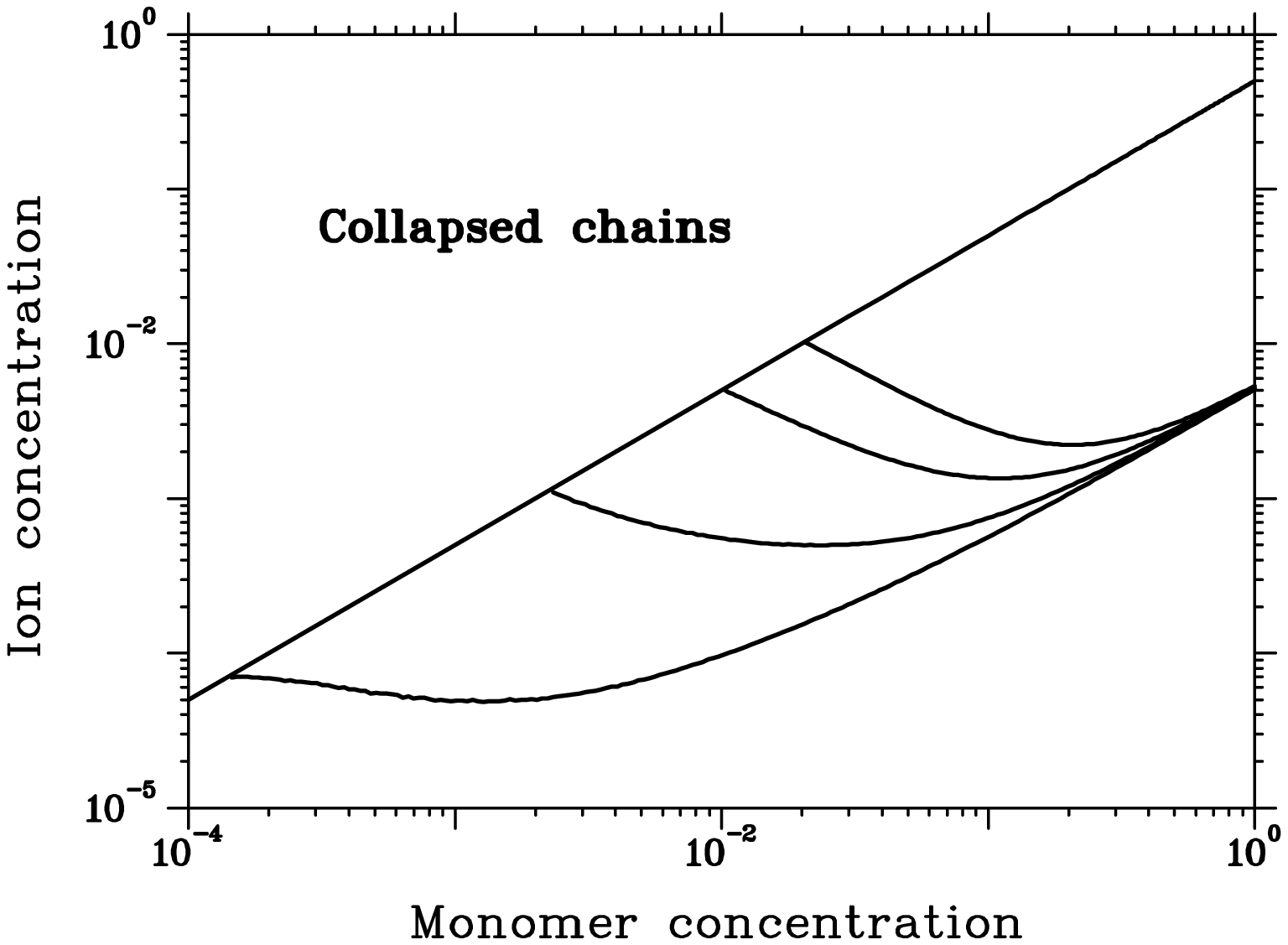}
\caption{Gelation lines for different values of parameters $l_B/b$
and $b/a$. From top to bottom $(l_B/b,b/a)$ = (2,1), (3,1),
(2,1.5), (3,1.5). Concentrations are given in $a^3$ units,
$N=100$.} \label{gel_lines}
\end{figure}

In Figure \ref{gel_lines} we show the gelation lines obtained for
different values of $l_B/b$ and $b/a$. Notice that the ion size
$a$ in $K_2$ in Eq. \eqref{K_2} is a parameter different than the
average distance between the neighboring monomers along the chain
$b$. It can be varied to account for the unknown size of the ions
in the crosslinks and the unknown dielectric constant of the local
medium around the links. For $a=b$ the gelation occurs at rather
high $\rho_i$ concentrations. This may explain why in solutions of
strongly charged polyelectrolytes, such as DNA and
Polystyrene-Sulphonate, where both metallic and organic
multivalent ions are hydrated around the chains ($a>b$), $z \geq
3$ are required to observe precipitation, and the gelation is not
observed at $\rho_i^* \simeq \rho_m/(Nz(z-1))$ even in semidilute
solutions but at much larger values of $\rho_i$. Instead in
polyelectrolytes with monomers with ionizable OH groups such as
acrylate groups, since metallic ions are dehydrated when forming
dicomplexes, divalent metallic ions do precipitate the chains, and
gels are formed in semidilute solutions\cite{Delsanti}.

We investigated the stability of the system to macroscopic phase
separation, and found no instabilities induced by electrostatics
below the gelation lines for the range of parameters used here (
$l_B/b \sim 2$, $\chi \leq 2$). Thus we conclude that there is no
segregation transition that competes with the gelation in the
regimes studied here.

In conclusion, we have determined the effect of long range
electrostatic interactions on gelation of polyelectrolyte chains
induced by multivalent ions. The approach can be used to describe
gelation of charged chains bridged by other type of linkers
through modification of $K_2$. We could also account for variable
weights of different types of crosslink functionalities $m$ by
allowing a $K_m$ dependence. It is of future interest to estimate
the effect of cyclization of the chains in the gelation theory of
polyelectrolytes.

We acknowledge the financial support of the NIH grant number
GM62109-02, the Institute for Bioengineering and Nanosciences in
Advanced Medicine (IBNAM) at Northwestern University and the NSF
grant number EEC-0118025.

\end{document}